# Collaborative systems and multiagent systems

**Assist.Prof. Alin Munteanu, PhD**
**Instructor Cristina Ofelia Sofran, PhD Candidate**
„Tibiscus" University of Timişoara, Romania

REZUMAT. Lucrarea prezintă câteva elemente de bază cu privire la domeniul sistemelor colaborative, un domeniu de maximă actualitate, şi cu privire la sistemele multiagent, dezvoltate în urma unui studiu temeinic asupra sistemelor cu un singur agent.

## 1. Collaborative as "jointly with"

In order to understand collaborative systems it is necessary to focus on those information systems which are intelligent partners, collaborating in solving problems. There is no doubt that an improvement of information systems would be appropriate. The use of the Internet has been increasing leading to better communication between people, no matter the place they find their selves. Still, many users are facing the limitations of the communication means that exist so far. The interfaces that can be directly manipulated are considered a way for any user to be able to access a computer. It is considered that anything the user needs is just a click away, there is no need of specifying the computer how to do a certain thing but only to ask what that thing is. But this is unfortunately a shallow truth, as for many applications the users don't have to write any code or commands, as they do when it comes to more complex problems. The user needs to "tell" the computer how to solve the problem. Although being complex tools, usual information systems are not aware of what the users are trying to do by using them, and are not of great help in solving problems. Information systems are supposed to solve problems without the user specifying each step to be executed.

To realize collaboration between the systems and their users, in order to obtain the needed information to be able to solve the problems of the users, it is imposed to use some artificial intelligence based technologies, referring not only to individual intelligent systems but to many intelligent systems that are working together, collaborating. This collaborative behavior is an important part of the intelligent behavior, and the collaborative behavior can only be understood after understanding the individual behavior. It is important to know how much of a behavior is collaborative and not coordinated or interactive. The distinction between collaborative and interactive is that interactive only refers to the action upon another thing or person, while collaborative refers to working jointly with the others.

In order to be able to build collaborative systems it is necessary to identify the capabilities that are to be added to the individual agents so that they become able to work with other agents. The modeling of collaborative has become a priority as during the last decade some important progress within the artificial intelligence field has been achieved.

## 2. Multiagent Systems within an intelligent environment

Multiagent Learning is at the intersection of multiagent systems and Machine Learning, two subdomains of artificial intelligence. Weiss has defined Multiagent Learning as "learning that is done by several agents and which is only possible due to the presence of several agents".

Traditional Machine Learning technologies usually imply a single agent that is trying to maximize some utility functions without having any knowledge about other agents within its environment. The multiagent systems refer to domains with many agents and there are considered some mechanisms for the interaction of independent agents' behavior.

The multiagent systems domain refers to the domains where several agents are involved and mechanisms for the independent agents' behaviours interaction have to be considered. The multiagent systems include any situation where an agent is learning to interact with other agents, even if the other agents have a static behavior. The main explanation for considering multiagent learning the situations when a single agent is actually learning is that its very learned behavior is being often used as a basis for a more complex and interactive behavior. Although a single agent is being involved in the learning process, its behavior is only obvious when other agents are present, and determines these other agents to participate to the collaborative and adversial learning situations. In the case when multiagent learning is happening by layering behaviors, all learning levels which imply interaction with other agents are actually contributing to and included in multiagent learning.

So far, most of the learning algorithms have been developed for a single agent. The learning of a single agent is focused on the way a single agent improves its own characteristics. Multiagent learning is out of question as long as the single agent learning does not affect or is not affected by a neighbor agent. Although an agent is not very aware of the other agents' existence, it will consider them as part of its environment and their behavior will be integrated in the learned hypothesis. Coordinating a group behavior seems to be possible by the learning of a single agent.

On the other side, the learning of a single agent will not always lead to best results within the multiagent learning to increase the efficiency. There is a difference of consciousness of other agents, and coordination, the question arising if higher level learning will implicitly lead to higher performances.

Distributed Artificial Intelligence (DAI) is a subdomain of artificial intelligence (AI), concerning systems that consist in several independent entities which interact in a certain domain. At the beginning, DAI was divided into two subdisciplines: distributed problem solving (DPS), focused on the management of information within systems that contain several subsystems working together for a certain goal and multiagent systems (MAS), referring to managing the behavior of a collection of independent entities or agents.

Multiagent systems (MAS) represent a subdomain of artificial intelligence which offer the principles for building complex systems that imply several agents and mechanisms for coordinating independent agents' behavior. As there is no generally accepted definition for the agent, it is considered that an agent is an entity, a sort of robot, with goals, actions and knowledge in a certain domain, located within a certain environment.

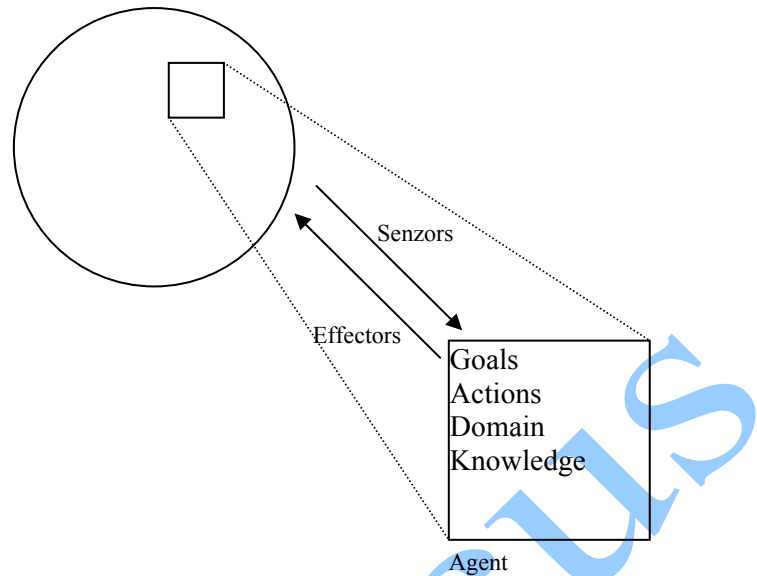

*Fig. 1:* *Single agent system*

Due to multiagent systems' complexity, there have to be found solutions for using Machine Learning technologies to manage this complexity. Before studying multiagent the alternative should be taken into consideration, the alternative being centralized systems with a single agent. The centralized systems have a single agent that is making all the decisions, while others are only helping. Single agents systems must be accepted as centralized systems from a domain that also allows the multiagent approach. A single agent system can have several input data and several actors. But if each entity is transmitting perceptions towards a single core process and receives actions from it, then there is only one agent represented by the core process. Within a single agent process, it is modeling itself, modeling the environment and its interactions with it.

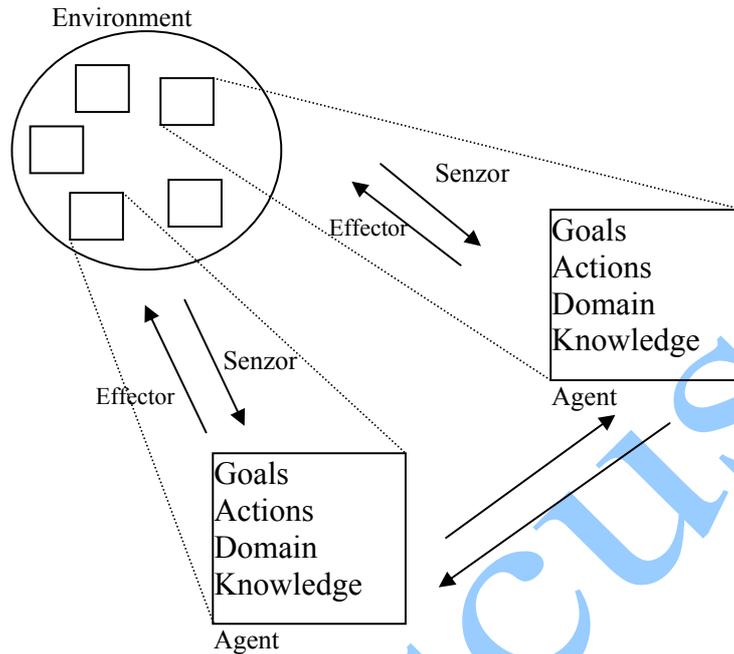

***Fig.2:*** *Multiagent system*

The multiagent systems are different from the single agent systems, and the difference consists in the existence of several agents which are modeling the goals and actions between themselves. Between the agents there is usually a direct interaction, but although this interaction is seen as stimuli of the environment, the communication between agents is distinct from the environment. In case of these systems, the dynamic of the environment can be defined by other agents.